# Managing Heterogeneous Substrate Resources by Mapping and Visualization Based on Software-Defined Network

Amir Javadpour[1], Guojun Wang[1], and Xiaofei Xing[1]
[1]School of Computer Science and Technology, Guangzhou University, Guangzhou, China, 510006
*Correspondence to: csgjwang@gzhu.edu.cn

*Abstract*— Network virtualization is a way to simultaneously run multiple heterogeneous architectures on a shared substrate. The main issue in virtualization of networks is the problem of mapping virtual networks to the substrate network. How to manage substrate resources when performing the mapping will have an effective role in improving the use of infrastructure resources and consequently better mapping. By writing a module in the controller for dynamic resource management, an initial mapping has been attempted until the request arrives, if sufficient resources are available, but until the arrival of the n request that their initial mapping is successful, writing the rules in the switches is postponed. The simulation results with the NS2 simulator showed that compared to the two similar approaches, the proposed method could reduce the delay and the cost by maintaining the acceptance rate.

## I. INTRODUCTION

With the development of networks, the need of using virtual networks for better performance and lower costs and generally resource management increased. Network virtualization is a promising technology for the future of the Internet in networks in which development and management is separated from service provision and multiple network services can run on a substrate network [1]–[5] and [6]. Whenever a new technology is introduced on the network, importance of its management is generally at its lowest level. As technology evolves and expands extensively, network management [2], [7]–[10]. In network virtualization, the primary entity is the virtual network[11]. The virtual network is a combination of active and passive network components (nodes and network links) on the substrate network. Virtual nodes connect each other through virtual links in the form of a [2], [12], [13]. By virtualizing resources, the node and link of substrate network can create a topology for multiple virtual networks with a wide range of features, and also host them in the same physical hardware. In addition, the abstraction introduced by the resource virtualization mechanism allows network operators to manage or modify the network in a dynamic with highly flexible way[14]–[16]. In order to improve virtualization, the software-based network has come to help virtual [1], [9], [17]–[19]. The software network is the core of the emerging network architecture in which the network control is split from the sending mechanism and can be programmed directly. In these networks, there is a central controller that has an attitude of network bandwidth and controls the sending of equipment (such as switches etc.) and can be configured by an interface [4], [12], [20]–[23] The software based network has three main parts which will be further discussed. Application: A part that uses the decomposed levels of control and data to achieve specific goals such as security mechanisms or network measurement methods. Control plane: This level manages the sender's equipment with a controller to achieve the specific purpose of the target. Data plane: A part of a shared protocol (Such as the OpenFlow protocol[24], [25]) that supports the control level and uses packets based on the configuration that is run on the controller. An important issue in network virtualization is the issue of virtual network embedding. By embedding virtual networks, every virtual network node will map to a network substrate node and each virtual network link will be mapped to one or more network substrate links. An important issue to consider is that substrate resources are consumed economically[5], [21]. So the mapping should be optimized. The problem of mapping in an optimal way is called the virtual network embedding problem. Another important issue is resource management which can be considered dynamically or statically. In the dynamic management after the mapping, there is a possibility to perform a remapping for better use of network resources. The following article will be presented as follows. In Section 2, a summary of related work and related studies will be discussed. In Section 3, the issue will be addressed and the existing challenges will be expressed. Then, the network model and how to calculate the required parameters will be expressed. In Section 4, the proposed method is expressed. Section 5 will simulate and evaluate the results.

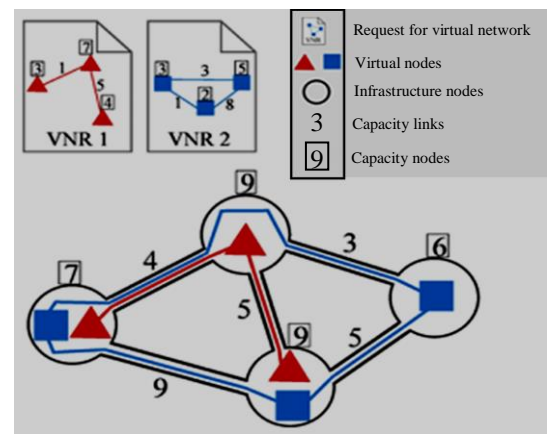

Fig. 1. suggested method for network virtualization.

## II. PREVIOUS RESEARCHES

The software-based network helped the virtualization of the networks by separating the logic of data control. Strategies for better management of resources in these networks are presented. Hoveydy et al [8] improved the performance by offering two approaches with and without the possibility of reconfiguring in a wide range of problem situations. In [26] Minjombi et al have provided a distributed, independent and artificial intelligence solution for resource allocation that the goal is to achieve self-configuration, self-optimizing, self-healing and self-protecting. The method proposed by Yo et al [27] considers the division of the substrate network for virtual link mapped onto multiple substrate paths. Also, the transfer route is intended to better use of substrate resources to increase the acceptance rate. In [28] Chaudhry et al, by proposing two new embedding algorithms for virtual network requesting, have achieved a better correlation between the node mapping phases and link mapping.

Trivisono in [21] presented a static and offline mixed integer linear programming formulation for a central controller to provide optimal calculation of end-to-end virtual paths to the substrate internet provider which considers multiple requests simultaneously. As a result, the controller can improve end-to-end streaming management. Mahmet et al [29] have proposed a VN mapping approach in software based networks aimed at balancing load on the substrate network and minimizing the delay of the controller to the switch. In [30] Feng et al proposed a method by combining link bandwidth allocations and flow tables for multiple control applications in a software based network. In addition, an integrated allocation model is built based on the amount of network resources in the software based network by introducing the amount for each resource which can provide fairly fair link bandwidth and minimum latency at a time. A flow planning policy was also proposed based on fairly broadband allocations that could achieve minimal overall latency. Minjombi et al [31] instead of allocating a fixed amount of resources for the entire life of VN, dynamically and optimistically allocate resources to nodes and virtual links dependent on previous needs. The simulation results show that in the dynamic approach, the virtual network acceptance rate will be better than the static approach. Then [31] Minjombi and colleagues proposed the proposed resource management system in comparison with the general software based network which includes a source manager and a database which has expanded into software based controller and has led to better management. In this approach, the switching resources and the link are synchronized. In the proposed method of this article has been tried to reduce remapping by using software based network and based on the method used in the previous article, [31] by keeping the rate of acceptance and considering the time window for the arrived requests. And ultimately reduce the delay caused by remapping and also reduces the cost.

## III. RESEARCH STATEMENT OF PROBLEM

In this section, we will first discuss embedding or mapping problem which is the most important topic in virtualization. The embedding problem is examined with an example in Figure 1 in which two virtual networks with three nodes on a network of substrate with four nodes are mapped. It's seen that a nod in the substrate network can host several virtual nods and substrate link can host multiple virtual links. As seen in the figure, one of the virtual links is mapped onto two substrate links and this suggests that a virtual resource can be a combination of several sources of substrate. Several issues should be considered in the mapping: firstly, the candidate substrate resources in mapping must support the needs of virtual resources. For example, a virtual link MBits 1000 and MBits100 cannot be mapped onto substrate link. Secondly, the CPU request of a virtual node should also be less than the capacity of the node substrate that it wants to map. If more needs to be done, more substrate resources should be reserved but the consumption of substrate resources should be cost-effective and therefore optimal mapping. Sometimes it is necessary to carry out some mapping for optimal mapping and acceptance of additional requests or in other word, mapped requests to be remapped so that more resources are freed up and the resources cannot be split up.

### A. Calculation of the parameters required for the proposed method

In this article, the proposed method is used to calculate the weight of the virtual links by Minjombi et al [31], that in the following, they will be listed in Table 2 below and outline them. Amount of using substrate link resources: For each virtual link $l_{u'v'}$ the parameter $R_{u'v'}$ represents the total amount of substrate resources used through the virtual link mapping above which is equal to the total amount of bandwidth used in the desired path that is fed by virtual currents passing through that virtual link as well as the total amount of memory resources used by the substrate switches to write those flow rules in the current tables in the switches.

TABLE I.   COMPARISON OF SOME PREVIOUS STUDIES AND RESEARCH STUDIES.

| Reference | Static dynamic | Goal |
|---|---|---|
| [32] 2006 | Relatively dynamic | Load balancing |
| [33] 2010 | dynamic | Distributed error-tolerant algorithm for mapping |
| [26] 2012 | static | Self- Configuration Self -Improvement |
| [27] 2008 | static | Increasing profits and reducing costs |
| [28] 2012 | static | Increasing acceptance and profit rates and reducing costs |
| [21] 2013 | static | Optimal calculation of end-to-end virtual paths |
| [29] 2014 | static | Fair allocation of link bandwidth and minimum total latency |
| [30] 2014 | dynamic | Increasing acceptance rates by guaranteeing the stability of service quality parameters |
| [31] 2014 | dynamic | Improving acceptance rates and reducing costs |

TABLE II. PARAMETERS REQUIRED TO CALCULATE THE WEIGHT OF EACH VIRTUAL LINK.

| Parameters | Explanation |
|---|---|
| $R_{u'v'}$ | The amount of link resources and switches used by the substrate |
| $A_{u'v'}$ | The amount of link sources and free and unchecked switches of substrate |
| $W_{u'v'}$ | Weight assigned to any virtual link |

The equation of the above definition is expressed as follows:

$$R_{u'v'} = \sum_{l_{uv} \in P_{l_{u'v'}}} (B_{u'v'}) + \sum_{u \in P_{l_{u'v'}}} (M_{u'v'})$$

Amount of unallocated resources of the substrate network: For each virtual link, the parameter $A_{u'v'}$, which represents the total amount of link resources and unallocated switch in the infrastructure network, has been computed so that for future mappings, resources are selected that have more load balances. This parameter is derived from the following relationship.

$$A_{u'v'} = \sum_{l_{uv} \in P_{l_{u'v'}}} (B_{uv} - B_{u'v'}) + \sum_{u \in P_{l_{u'v'}}} (M_{uv} - M_{u'v'})$$

Weight of virtual links: By calculating the above parameters, it is easy to calculate the amount of load on any virtual link calculated by the parameter $W_{u'v'}$ by reducing the total weight of resources from the total available resources weight. Then, by sorting out the calculated weight of virtual links in a downward spiral, it would prioritize a link with more weight to reduce the burden imposed on the substrate network.

$$W_{u'v'} = R_{u'v'} - A_{u'v'}$$

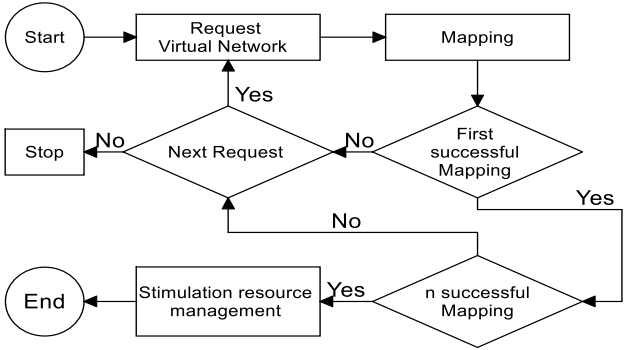

Fig. 2. suggested method for network virtualization.

### B. Proposed algorithm

In the proposed method, the process of virtualization of the network will be performed and controlled by a software based network. In the software network, a module for controlling resources is written in the controller that by stimulating it, the specified rules will be written in the database intended for the network. Then, according to the database, only the modified rules in the corresponding switches are added and thus the continuous updating of all switches will be prevented. The proposed algorithm is first to wait for the virtual network request to arrive, then each mapping request is executed. If sufficient resources are available to respond to the request, a successful initial mapping will be done otherwise the request will be rejected and awaiting further requests. Requests whose initial mapping has been successful will put in a queue at the request to make the n initial mapping successful. By performing the nth early successful mapping, the management module in the stimulated controller and the parameter of mapped virtual links' weight calculated in the same order as previously stated and will arrange them in descending order. The difference between this method and the method of Minjombi et al. [31] is the addition of a time window for successful early mappings and after each successful mapping, the flow rules are not written in the switches, but they are waiting in the mapping queue and no rule will be written in any switch before reaching the n mapping. The controller is triggered with the nth successful mapping, and then all the rules of the n successful mapping flow are written together in the switches. Flowchart of the proposed method is shown in Figure 2. The weights obtained for each virtual link have been recalculated and arranged by changing the mode of the network substrate resources, such as mapping or remapping or retrieving resources after the expiration of a request, to provide the conditions for remapping, for better use of resources, specify the priority of the links to be remapped. Regarding the changes that have taken place, stream rules are only written or deleted or changed in the related switches.

### IV. EVALUATION

In this article, the NS2 emulator version 2.35 has been used, which has been simulated by adding two modules 16 of the SDN network. In all the simulation of the link speed in sending Gb/s 1, at the bottleneck position 1Gb/s, the minimum trip time of the RTT (Trip Round Time) information from s 100m and the TCP protocol was used. In this simulation, a Dumbell tree topology is considered with a bottleneck at origin, and the simulation is run for a 30 second period. The bottleneck buffer size of the link is greater than the latency bandwidth result in all cases (100 packets), the size of the IP data packet is considered 1500 bytes. In order to evaluate the design of a floodlight controller, it has been expanded by writing a management module to execute the desired functions of the design. Mapping the virtual network nodes into substrate network was carried out by implementing a greedy approach and mapping virtual links substrate links with multi-commodity flow and without splitting the path. The memory capacity of the switches and the bandwidth of the links are also considered with a uniform distribution of between 100 and 250 units. The number of switches is by default 14 switches. In this article, it is assumed that each current rule consumes 1 unit of the memory of the switch. The number of virtual network requests in simulation is up to 1500 requests. In order to illustrate the results of the article, the results compared with two other approaches. In the SSPSM approach [27], a time window is used to receive requests and the split path and link transfer are used for mapping. The SDN-VN approach [31], which was also based on the work of this research, has been using the SDN network to manage and control resources and requests. While the proposed method in this article, in addition to this, time window included for more effective initial mapping and therefore less mapping. Figure 3 shows the acceptance rate for all three articles. As can be seen, the rate of acceptance has significantly increased compared to the SSPSM approach, compared to the SDN-VN approach, however slight but it has increased and did not decrease. This increase is due to the possibility of remapping in the SDN-VN method and the proposed method of this study can be expressed. In Figures 4 and 5, the average use of link sources and switches is compared in three approaches. These parameters indicate that, on average, how much has been used in relation to the total available resources and express the average pressure on the sources. It is seen that initially, the link load in the SSPSM is lower than in the other two, but with increasing number of requests, this amount increases, while the increase in the number of requests in the other two approaches leads to a decrease in the link load, this is due to the static management in SSPSM and dynamic management that can be expressed in the two approaches mentioned above. This difference in management method is also visible in the amount of use of switching resources. In Figure 6, the network latency parameter is compared to three papers. The network latency has dropped by increasing the number of switches. Due to considering the window of time in the proposed method of this study, this reduction rate can be expressed. In Figure 7, the mapping cost is compared. The cost of mapping by dividing relation (1), which represents the total amount of resources used in virtual request mapping, is derived from the number of mappings. It is seen that the cost parameter in the proposed method of this article has been reduced compared to the above two approaches. From this chart, and chart 5 and 6, and according to the definition of cost, it can be said that the cost reduction, while the use of resources has increased, is the result of an increase in the number of records.

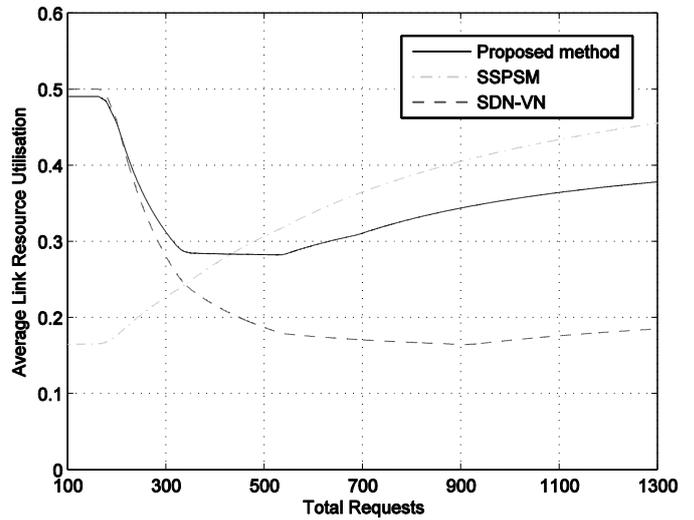

Fig. 4. Results of the evaluation of the average consumption of resources of the network infrastructure link.

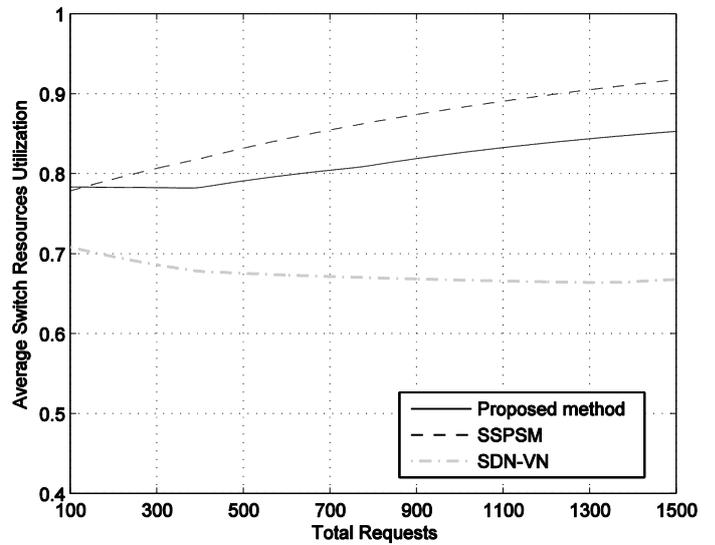

Fig. 5. Results of the evaluation of the average consumption of resources of the infrastructure network switch.

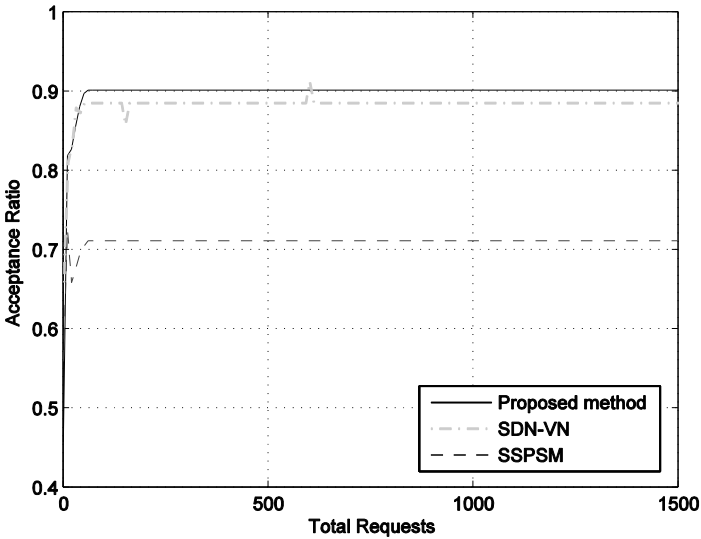

Fig. 3. Compare admission rates on the network.

## V. CONCLUSION

What is outlined in this article is to provide a dynamic management approach using a software based network to better map the requests of virtual networks to the substrate network. In this method, using a resource management module in the controller, which is actually a mode-operation machine, possibility of remapping for requests was provided. Also, with the controller, only the modified rules in the corresponding switches will be changed and the need for the continuous update of all network switches will be lost. In this study, for the final mapping, time window is considered, which has led to a reduction in cost and a reduction in network latency which can be said that the cost reduction has been due to the increasing number of mappings.

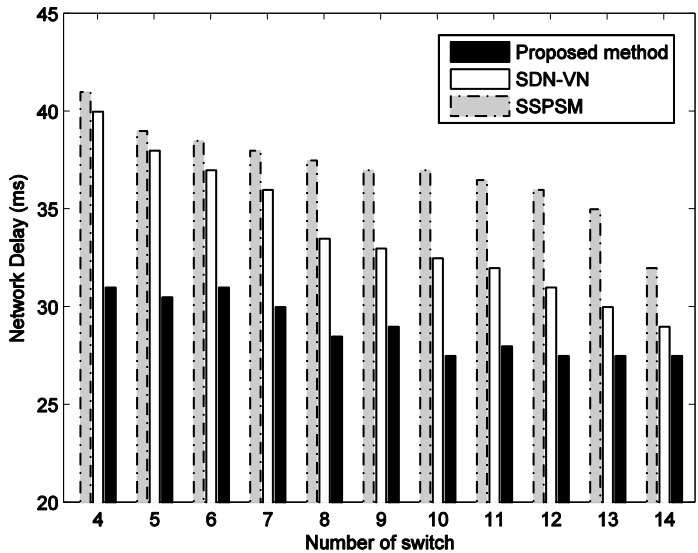

Fig. 6. Results of network latency evaluation.

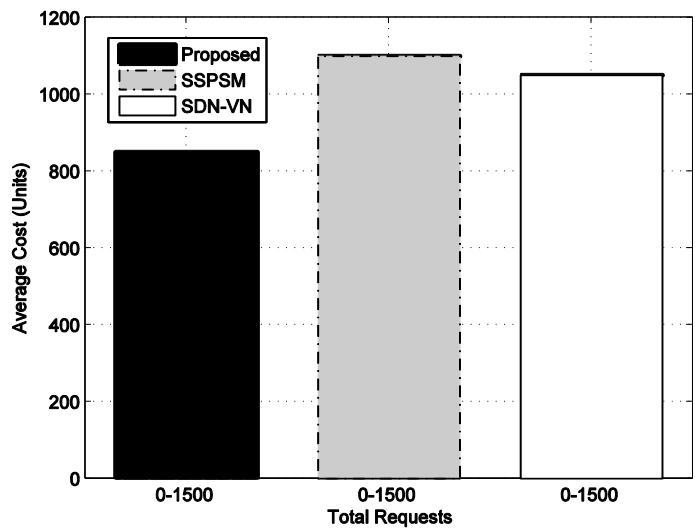

Fig. 7. The average cost in the network after virtualization with each of the approaches.

ACKNOWLEDGMENTS

This work is supported in part by the National Natural Science Foundation of China under Grants 61632009 & 61472451, in part by the Guangdong Provincial Natural Science Foundation under Grant 2017A030308006 & 2016A030313540, Hgh-Level Talents Program of Higher Education in Guangdong Province under Grant 2016ZJ01 and Guangzhou Science and Technology Program under Grant No. 201707010284.